\documentstyle[aps,twocolumn,graphicx]{revtex}
\begin{document}
\draft
\twocolumn[
\hsize\textwidth\columnwidth\hsize\csname@twocolumnfalse\endcsname
\title{Luminosities of High-Redshift Objects in an Accelerating Universe}
\author{Daniel Enstr\"{o}m, Sverker Fredriksson and Johan Hansson}
\address{Department of Physics, Lule\aa \ University of Technology,
  SE-971 87 Lule\aa, Sweden}
\date{\today}
\maketitle
\begin{abstract}
The results from the Supernova
Cosmology Project indicate a relation
between cosmic distance and redshift that
corresponds to an accelerating Universe,
and, as a consequence, the presence of
an energy component with negative pressure.
This necessitates a re-evaluation of
such astrophysical luminosities that have been
derived through conventional redshift analyses
of, {\it e.g.}, gamma-ray bursts and quasars.
We have calculated corrected luminosity
distances within two scenarios; the standard
one with a non-zero cosmological constant,
and the more recently proposed ``quintessence'',
with a slowly evolving energy-density component.
We find luminosity corrections from
$+30$ to $-40$ per cent for redshifts
with $z=0 - 10$.
This finding implicates that the SCP data do
not, by themselves, require a revision
of the current, rather qualitative modeling
of gamma-ray bursts and quasar properties.
\end{abstract}
]
\narrowtext

\begin{figure}
\includegraphics[scale=0.5]{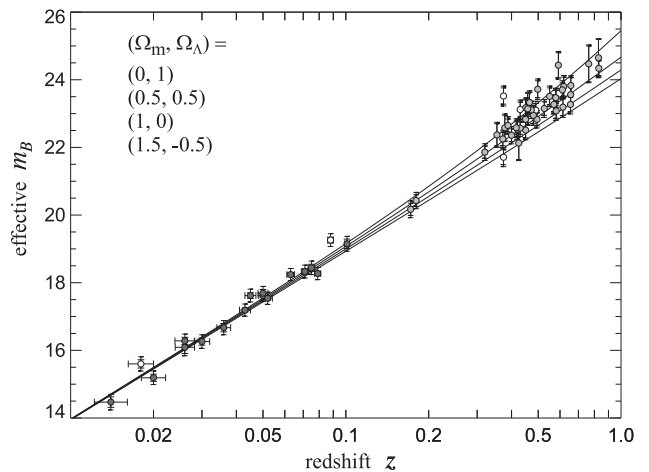}
\caption{Effective magnitude, $m_B$, vs. redshift,
$z$, for 42 high-redshift and 18 low-redshift
Type Ia supernovae. The solid curves are the expected
relations for a flat Universe with
$\Omega_{m}+\Omega_\Lambda=1$, and for a choice
of combinations, given in the graph in
the same vertical order as the curves.
The plot is taken from \protect\cite{perlmutter98b},
but with an edited layout. The low-redshift data are taken
from the Calan/Tololo survey \protect\cite{hamuy96}.}
\label{SCP_fig}
\end{figure}

In 1938, Baade \cite{baade38}
suggested supernovae as ``standard candles''
for measuring various cosmological parameters.
At closer distances they should reveal the
Hubble constant, while at higher redshifts
they were assumed to eventually indicate a
universal deceleration \cite{tammann79,colgate79}.
Measurements of the Hubble constant became
feasible in the 1980s, while the attempts
to detect a universal deceleration failed, due
to a lack of observable high-redshift supernovae.
When the Supernova Cosmology Project (SCP)
was initiated in 1988 its primary goal was
to determine cosmological parameters
through the magnitude-redshift relation of
Type Ia supernovae. Goobar and Perlmutter
\cite{goobar95} showed that by studying
this relation one might be able to separate
the relative contributions to the density of
the Universe into one part, $\Omega_m$, due to masses
(including the hypothetical ``dark matter''),
and another part, $\Omega_\Lambda$, due
to a non-zero value of the cosmological constant,
$\Lambda$, as given by the Einstein equations.
The latter is looked upon as a density
of ``dark'' energy hidden within the physical
vacuum. As of March 1998, more than 75 supernovae
of Type Ia at redshifts $z=0.18-0.86$ had been
discovered and analyzed
\cite{perlmutter95b,perlmutter97a,perlmutter97b,perlmutter97c,perlmutter97d,perlmutter98a}.
The results are summarized in Fig. \ref{SCP_fig}.
A similar study by the ``High-z
Supernova Search Team''\cite{riess98}
has produced results in agreement with those
of the SCP.

A deviation from the expected magnitude-distance
relation is seen. Assuming a flat universe, as
predicted by the hypothesis of inflation
within the standard Big Bang scenario
\cite{guth81}, it is clear that the
major energy density must be of the
``vacuum'' type. This finding obviously
implicates that a re-analysis of
astrophysical data (and possibly
of theoretical models) deduced from
cosmologies with $\Omega=\Omega_m$
is necessary. Examples of cosmic phenomena
that need to be reconsidered are gamma-ray
bursts (GRBs), where redshifts have been
found in apparent host galaxies,
as well as quasars and active galactic nuclei.
Since one of the great mysteries of the GRBs
is the enormous energy
release in the form of gamma photons, it is
important to estimate the
corrections implied by the SCP results.
\noindent

The luminosity, ${\cal L}$, of high-redshift
objects, such as GRBs, are determined
using the luminosity distance, $d_{\cal L}$,
and the flux on the detector, $\phi$
(in erg s$^{-1}$ cm$^{-2}$), through the relation
\begin{equation}
\label{energyflux}
{\cal L}=4\pi d_{\cal L}^2 \phi,
\end{equation}
assuming a spherically symmetric
energy outflow. In this Letter,
we examine the implications of
the SCP results for $d_{\cal L}$ and hence
also for ${\cal L}$. There
are two different approaches that have raised
a particular interest in the current literature.
The first one builds on a traditional use
of the cosmological constant, $\Lambda$,
as first suggested by Einstein in a
different context. The other one includes
a recent proposal of an additional
energy-density component, parametrized as a slowly
evolving scalar field, $\varphi$, with a
positive potential energy \cite{caldwell98}.
This so-called quintessence (see \cite{wang99},
and references herein) is a dynamical, spatially
inhomogeneous, energy, resulting in a
negative pressure. Unlike the cosmological constant,
this scalar field slowly changes its
contribution to the energy density of the universe,
not only due to the expansion, but also
through its slow approach toward a lower potential
energy. The equation of state, {\it i.e.},
the relation between pressure, $p$, and
density, $\rho$, for this
energy component is parametrized as
$p=w\rho$, where the contant $w\in(-1,0]$.
The case $w=-1$ corresponds to a nonzero
cosmological constant. In \cite{wang99},
a fit was made to a wealth of cosmological data,
resulting in $w\approx -0.65 \pm 0.07$.
This is well in line with the limits placed on
$w$ by the SCP (see Fig. 10 in \cite{perlmutter98b}).

The distances to high-redshift objects have
conventionally been estimated within
a so-called Friedmann-Robertson-Walker (FRW)
cosmology with $\Omega=\Omega_m$.
In light of the recent SCP results,
these assumptions have to be modified,
and the analysis becomes a bit more
complicated. The result stated by the
SCP group \cite{perlmutter98b} is
$\Omega_m^{flat}=0.28^{+0.09}_{-0.08}\,^{+0.05}_{-0.04}$
for a flat Universe, defined by $\Omega=1$.
Hence, roughly $70$ per cent of the energy density
is in the ``vacuum'' form. This energy acts
as an effective repulsive potential in the
Friedmann equation, making the universe
expand at an ever increasing speed,
and the SPC \cite{perlmutter98b} states
that the data are in line {\it only}
with a currently accelerating Universe.
Nevertheless, this statement is limited to
the redshift range of the studied supernovae,
{\it i.e.}, out to $z\approx 1$. We therefore
assume that the FRW cosmology used by the SCP when
fitting the data is valid also at higher redshifts,
where we apply the two different approaches
mentioned above. The basic Friedmann
equation, neglecting a radiation energy density,
can be written as
\begin{equation}
\label{friedmann_eq}
H^2=\left(\frac{\dot{a}^2}{a^2}\right)=
\frac{8\pi G}{3}\left(\rho_m+\rho_\Lambda+\rho_\varphi\right)
-\frac{k}{a^2}.
\end{equation}
Here $a=a(t)$ is the spatial scale factor in the
FRW metric, $G$ is Newton's gravitational constant,
$k$ is the Riemannian curvature parameter, and
$\rho_m$ is the matter density. The vacuum-energy
and quintessence densities,
$\rho_\Lambda$ and $\rho_\varphi$, are defined as
\begin{eqnarray}
\rho_\Lambda & = & \frac{\Lambda}{8\pi G}, \\
\rho_\varphi & = & \frac 12 \dot{\varphi}^2+V(\varphi),
\end{eqnarray}
where $\Lambda$ is the cosmological constant,
while $\varphi$ and $V(\varphi)$ are the field
and potential energy in the quintessence model.

The various contributions to the critical
density $\Omega$ from $\rho_m$, $\rho_\Lambda$
and $\rho_\varphi$, as well as from
the curvature term $\-k/a ^2$, are given by
\begin{eqnarray}
\label{critdens_def}
\Omega_m & = & \frac{8\pi G}{3H_0^2}\rho_0, \\
\Omega_\Lambda & = & \frac{\Lambda}{3H_0^2},\\
\Omega_\varphi & = & \frac{8\pi G}{3H_0^2}\rho_{\varphi_0},\\
\Omega_k & = & \frac{-k}{a_0^2H_0^2},
\end{eqnarray}
where subscript ``0'' stands for the current
($t=t_{0}$) value of each quantity, including that
of the Hubble constant, $H_{0}$.
The sum is fixed by
$\Omega_m+\Omega_{\Lambda}+\Omega_{\varphi}+\Omega_k=1$
for all cosmologies. Using the scaling relations
\cite{bergstrom99} $\rho_m\propto 1/a^3$
and $\rho_\varphi\propto 1/a^{3(1+w)}$,
we reformulate (\ref{friedmann_eq}) as
\begin{eqnarray}
\label{friedmann3}
H^2/H_0^2 & = & \Omega_m(1+z)^{3}+\Omega_\Lambda+
\Omega_k(1+z)^{2} \nonumber \\
 & + & \Omega_\varphi (1+z)^{3(1+w)},
\end{eqnarray}
where we have used the definition of redshift,
\begin{equation}
\label{z_def}
1+z=\frac{a_0}{a}.
\end{equation}
The luminosity distance, $d_{\cal L}$, is defined by
\begin{equation}
\label{d_L}
d_{\cal L}=a_0r_1(1+z),
\end{equation}
where $r_1$ is the comoving distance traveled by a
photon emitted from the source at time $t=t_1$.
The quantities $a$, $r$ and $t$ are related by the
equation of a radial, lightlike geodesic of the FRW metric,
\begin{eqnarray}
\label{r1_def}
\frac{dr}{dt}  =  \frac{\sqrt{1-kr^2}}{a(t)}
\Rightarrow \nonumber \\
\int_{0}^{r_1}\frac{dr}{\sqrt{1-kr^2}}
 =  \int_{t_1}^{t_0}\frac{dt}{a(t)}.
\end{eqnarray}
The relationship
\begin{equation}
\label{H_z_rel}
H=\frac{d}{dt}\left[log\left(\frac{a}{a_0}\right)\right]
=\frac{-1}{1+z}\frac{dz}{dt},
\end{equation}
can be used to transform the time integral in
Eq. (\ref{r1_def}) to an integral over $z$, as
\begin{equation}
\label{dt_trans}
\int_{t_1}^{t_0}\frac{dt}{a(t)} =
\int_{0}^{z}\frac{dz'}{\sqrt{g(z')}},
\end{equation}
where $g(z)$ is the expression in the rhs
of Eq. (\ref{friedmann3}). The integral over $r$
in Eq. (\ref{r1_def}) has the solutions
\begin{eqnarray}
\label{r1_sol}
\left\{
\begin{array}{l}
\frac{arcsin(\sqrt{k}r_1)}{\sqrt{k}} \hspace{0.25cm} (k>0)\\
\\
r_1\hspace{0.25cm} (k=0)\\
\\
\frac{arcsinh(\sqrt{-k}r_1)}{\sqrt{-k}}\hspace{0.25cm} (k<0)
\end{array}
\right.
\end{eqnarray}
Combining Eqs. (\ref{r1_def}), (\ref{dt_trans})
and (\ref{r1_sol}) leads to an expression for
$r_1$ as a function of $z$, given by
\begin{equation}
\label{r1_z}
r_1=\frac{1}{ \sqrt{ \left| -\Omega_k \right| } }
S\left\{\sqrt{\left|-\Omega_k \right| }
\int_0^{z}\frac{dz'}{\sqrt{g(z')}}\right\},
\end{equation}
where $S\left\{x\right\}$ takes on the
forms $sin\left\{x\right\},\,x,\,sinh\left\{x\right\}$
for the three different curvatures given by $k=+1,0,-1$,
{\it i.e}, for a closed, flat and open Universe.
The final expression for $d_{\cal L}$ then becomes
\begin{equation}
\label{dL_z}
d_{\cal L}(z)=a_0(1+z)\frac{1}{ \sqrt{ \left| -\Omega_k \right| } }
S\left\{\sqrt{\left|-\Omega_k \right| }
\int_0^{z}\frac{dz'}{\sqrt{g(z')}}\right\}.
\end{equation}
We compare the results from Eq. (\ref{dL_z}) with
the standard expression for the luminosity distance in an
FRW universe with $\Omega=\Omega_m$, {\it i.e.}, with
\begin{equation}
\label{dL_analytic}
d_{\cal L}^0(z)=\frac{1}{H_0q_0^2}\left[ q_0z+
\left(q_0-1\right)\left(\sqrt{1+2zq_0}-1\right)\right],
\end{equation}
where $q_0=\Omega_m/2$, since $\Lambda=0$ in this case.
For simplicity we have used $\Omega_m=0.28$ in all
calculations, since this is a result in good
agreement with the SCP and other observations
\cite{bahcall98,daly98}. The results are quantified as
$\alpha$, the squared ratio between the corrected
$d_{\cal L}$ and the traditional $d_{\cal L}^0$.
According to Eq. (\ref{energyflux}) this is
also equal to the ratio between the corrected energy
outflows (or luminosities) and the ``published''
ones (assuming that $\Omega_m=0.28$ has
been used). Hence,
\begin{equation}
\label{e_corr}
\alpha=\frac{E_{corr}}{E}=\left(\frac{d_{\cal L}}{d_{\cal L}^0}\right)^2.
\end{equation}

\begin{figure}[h]
\includegraphics[scale=0.45]{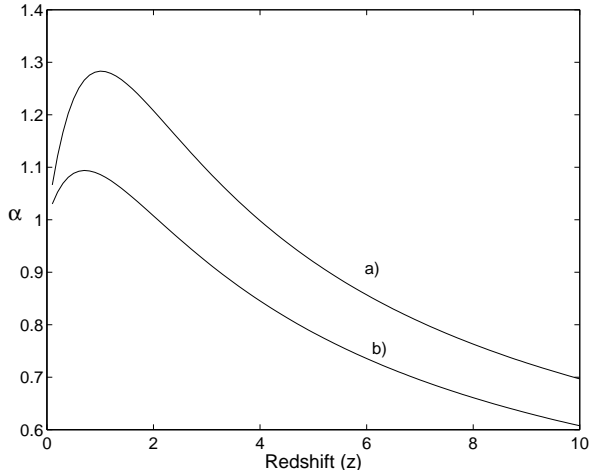}
\caption{The luminosity correction factor, $\alpha$,
as a function of redshift, $z$, for the ``conventional''
vacuum-energy and the quintessence scenarios. Curve a) has
$\left(\Omega_m,\Omega_\Lambda,\Omega_\varphi\right)$
= $(0.28,0.72,0)$ and curve b) has
$\left(\Omega_m,\Omega_\Lambda,\Omega_\varphi\right)$
= $(0.28,0,0.72)$ and $w=-0.65$.}
\label{fig_quint}
\end{figure}

The results for both scenarios
(``conventional'' flat Universe, and quintessence)
are shown in Fig. \ref{fig_quint}.
There is a clear positive correction for
low $z$ values, although only in the $10-30$
per cent range. Such an enhanced, corrected luminosity
has been intuitively expected by some groups
for GRBs at those ``low'' redshifts \cite{kulkarni99}.
It is therefore comforting that the correction is
so small, which means that most qualitative
conclusions about energy flows, drawn from the
published values of GRB redshifts, remain unchanged.

For higher $z$ values, luminosities must be corrected
{\it downward}. In the first scenario,
this qualitative difference between
low and high $z$ values has to do with the fact
that the presence of $\Omega_{\Lambda}$ influences
the development of the Universe in two different ways.
First, it contributes an enhanced
energy density that reduces the negative curvature
of the Universe, and
second, it provides a negative pressure that
accelerates the expansion. At {\it low} $z$,
{\it i.e.}, for observations in our vicinity,
the reduced, negative curvature of the open cosmology
in the ``denominator'' of Eq. (\ref{e_corr})
is negligible, since the Universe is
approximately flat in our neighborhood.
The effect of the vacuum energy in the
``numerator'' is therefore dominating,
which explains the positive correction.
At {\it high} $z$, the opposite is valid,
{\it i.e.}, the effect due to the difference
in curvature dominates, and the correction due
to the repulsion is negligible.
If the  vacuum energy is enhanced beyond
that of a flat Universe, {\it i.e.}, so that
$\Omega > 1$, the repulsive effect dominates the
correction out to even higher $z$ values.
Also, the maximal correction at $z \approx 1.5$
grows rapidly with increasing $\Omega_{\Lambda}$.
In a hypothetical Universe with
$\left(\Omega_m,\,\Omega_\Lambda\right)=(0,1)$
it reaches a factor of about two.

In the quintessence scenario, the trends in
Fig. \ref{fig_quint} have the same origin as in
the ``conventional'' case. The scalar field $\varphi$
has a repulsive effect, just as the cosmological
constant $\Lambda$, and affects $a(t)$ in the
same way. It should be noted that $\varphi$ is a
function of time, and it is not obvious that $w$
is a constant. However, it is argued in
\cite{wang99} that the physical, observable
consequences of a time-varying $w$
are negligible.

In conclusion, the luminosity correction
in the redshift range of ``identified'' gamma-ray
bursts, such as GRB990123 \cite{feroci99},
is $10-30$ per cent, depending on the cosmological
scenario. For a typical quasar at redshift $z\sim 5$,
the correction is negative, giving a luminosity
$80-90$ per cent of the one estimated for a
Universe with $\Omega=\Omega_m=0.28$.
The main result of our study is that current models for
luminous objects at high redshifts do not need to be
qualitatively altered due to the SCP supernova
results.

\end{document}